\documentstyle[preprint,revtex,eqsecnum,oldlfont]{aps}
\tightenlines
\begin{document}
\draft
\preprint{DFTT 7/93, February 1993}
\begin{title}
The quasi-elastic nuclear response
\end{title}
\author{A. De Pace}
\begin{instit}
Istituto Nazionale di Fisica Nucleare, Sezione di Torino,
  I-10125 Torino, Italy
\end{instit}
\author{M. Viviani}
\begin{instit}
Istituto Nazionale di Fisica Nucleare, Sezione di Pisa,
  I-56100 Pisa, Italy
\end{instit}
\begin{abstract}
We explore the nuclear responses at intermediate energies,
particularly in the
spin longitudinal $\sigma\cdot q$ and spin transverse
$\sigma\times q$ isovector channels, within the continuum
random phase approximation framework.
We also employ an extension of the standard random phase
approximation to account for the spreading width of the
single particle states through the
inclusion of a complex and energy-dependent nucleon self-energy.
The nuclear responses are then used as the basic ingredient to
calculate hadronic reactions in the Glauber theory framework.
Here both one and two-step contributions to the multiple scattering
series in the quasi-elastic peak region are taken into account.
We find evidence for shell effects in the one-step response and a
strong dependence on the momentum regime of the two-step
contribution.
\end{abstract}
\pacs{PACS numbers: 25.40.Kv, 25.40.Ep, 25.30.Fj, 24.10.Cn}
\narrowtext
\section{INTRODUCTION}
\label{sec:intro}

The interest in the nuclear responses at intermediate energies,
in particular in the quasi-elastic peak (QEP) region, is still
widespread: indeed, the body
of experimental data presently available in this energy regime is
not easily amenable to a unified and consistent understanding.

For example, the maxima of the (p,n) cross-sections \cite{tadd91}
appear hardened, over a wide range of momentum transfers, by about
25~MeV with respect to the ones observed in the (p,p$'$) reaction
\cite{chrien80}, an outcome still waiting for a convincing
interpretation in terms of either many-body or reaction-mechanism
effects.

The issue is further complicated by the experiments with heavier
probes: in fact, the hardening of the QEP position found in the
($^3$He,t) \cite{berg87} and in the (d,2p) \cite{sams91} reactions
at low transferred momenta turns into a softening when the momentum
transfer is increased.
Although this behaviour might be related to the composite nature
of the probes, it is notable that the same effect it is observed
also in pion single charge-exchange reactions \cite{peter92}.
On the other hand, the predicted hardening of the isovector
transverse {\boldmath $\sigma$}$\times{\bf q}$ response appears well
established according
to the available data of deep inelastic (e,e$'$) scattering
\cite{meziani85}.

In this paper, we shall mainly concentrate on the
{\boldmath $\sigma$}$\cdot{\bf q}$ and
{\boldmath $\sigma$}$\times{\bf q}$ isovector channels.
Actually, some  evaluations of the spin-isospin ($\sigma\tau$)
responses have already been performed in the framework of the
bound state \cite{alb87,toki87,alb88,adp90,adp91} or
continuum \cite{arima88,ichi89} random phase approximation (RPA)
with the conventional particle-hole (ph) interaction for the
$\sigma\tau$ channel.
As it is well known, the latter includes, beside the Landau-Migdal
parameter $g'$ (or, equivalently, a short range interaction of some
sort), the exchange of the pion and the rho meson.

A suitable test for these calculations is represented by the above
mentioned data  of deep inelastic electron scattering, where the
{\boldmath $\sigma$}$\times{\bf q}$ response has been separated out
in a few nuclei over a wide range of momentum transfers.
Here the virtual photon is exploring the whole of the nucleus,
which thus has all its constituents responding to the probe.

When, however, one comes to consider strongly interacting probes,
the ones offering a real hope of disentangling the elusive
{\boldmath $\sigma$}$\cdot{\bf q}$ response, the additional problem
is faced of appropriately dealing with the reaction mechanism, or,
in other words, with the distortion of the impinging and outgoing
hadronic waves.
This is essential, since here the response of the nucleus is
mostly confined to the surface region and one has to accurately
assess how much of the latter is actually involved in the process.

In this connection, the authors of Ref.~\cite{ichi89} resorted to
the Distorted Wave Born Approximation (DWBA), whereas in
Refs.~\cite{adp90,adp91} the Glauber theory \cite{glauber59} in
the one-step approximation was employed.
The former approach, as it is well-known, is fully
quantum-mechanical; however, in the high energy limit and for
nearly forward scattering the DWBA practically coincides with the
eikonal approximation and thus is equivalent to the Glauber
theory: but the latter allows one to deal with multistep processes,
which are likely to give a substantial contribution at large
momentum and energy transfers.

This is indeed what happens and we shall explore the contribution
the two-step processes provide, particularly in charge-exchange
reactions, where their importance is expected to be greater.

Turning to the many-body aspect of the problem, we calculate
the $\sigma\tau$ nuclear responses in the continuum RPA with the
($g'+\pi+\rho$) ph interaction, but utilizing a method different from
the one originally proposed by Shlomo and Bertsch
\cite{bertsch75}, which has been widely exploited in this kind of
calculations.
Our approach is entirely worked out in momentum rather than in
coordinate space and allows one to incorporate in a natural way
the spreading width of the ph states.

In fact, continuum RPA naturally accounts for the escape width of
the particle states, ignored, instead, in a harmonic oscillator
basis. However, the spreading width of the states is also quite
important for a realistic description of the $\sigma\tau$ responses.
The inclusion of the latter is achieved in Ref.~\cite{ichi89}
through the direct introduction of an optical potential in the
Schroedinger equation for the single particle (but not for the hole)
states. The disadvantage of such an approach lies in the ensuing
lack of orthogonality between the single particle wave functions
that needs then to be cured with a rather cumbersome procedure
\cite{ichi91}.

Alternatively, Smith and Wambach \cite{wambach88} introduced a
phenomenological complex and energy-dependent self-energy that
directly couples the ph to the 2p2h sector of the Hilbert space,
a procedure that should simulate the much more computer-time
consuming Second RPA (SRPA) \cite{yanno83}.
We adopt here an analogous approach: however, instead of
introducing the coupling at the level of the RPA eigenstates we do
so at the level of the bare ph states and {\it then} proceed to
the construction of the RPA solution.
The two approaches differ since the validity of the former requires
the near diagonality and state independence of the ``collective''
self-energy in the RPA basis, whereas in the latter the ph
self-energy in the ph basis is required to be almost diagonal.
Notably, the two frameworks lead to quite similar results, as far
as the nuclear response is concerned.

In this paper we shall test the above outlined model against deep
inelastic electron scattering data ({\it volume responses}) on
$^{40}$Ca as obtained in Saclay \cite{meziani85}.
For the {\it surface responses} we shall analyze the
data obtained with the charge exchange (p,n) reaction at Los Alamos
\cite{tadd91}.
We shall consider as well the (p,p$'$) data \cite{chrien80}.
In both cases the proton probe has an energy of 795 MeV.
This set of data represents a large body of what is nowadays
available on the experimental side.
We do not analyze here the (d,2p) and ($^3$He,t) data
(see, however, Ref.~\cite{adp91}): the internal structure of
composite probes would require an appropriate treatment, since it
has a strong impact on how the probe's and target's spins are
coupled \cite{bugg87}, and might also qualitatively influence the
distortion effects \cite{dmitr92}.

We organize the paper as follows: in Sec.~\ref{sec:vol} we present
our formalism of the continuum RPA and the way we deal with the
spreading width problem; in Sec.~\ref{sec:surf} we discuss the one
and two-step contributions to the surface response within the
Glauber theory, taking into proper account the cylindrical geometry
of the reaction.

In Sec.~\ref{sec:results} we show how the various ingredients of
the model affect the response functions and then test the model
against the already mentioned body of experimental data.

Finally, in the concluding Section we summarize our results.

\section{Volume Response Functions}
\label{sec:vol}

\subsection{The Polarization Propagator and the Continuum RPA}
\label{subsec:polprop}

As it is well-known, the nuclear response function to an external
probe is obtained through the imaginary part of the polarization
propagator \cite{walecka71} (non-diagonal in momentum space for a
finite system), which reads
\FL
\begin{eqnarray}
  \Pi({\bf q},{\bf q}';\omega) &=&
    \sum_{n\ne0}<\psi_0|{\hat O}({\bf q})|\psi_n>
    <\psi_n|{\hat O}^\dagger({\bf q}')|\psi_0>\nonumber\\
  && \times\left[{1\over\hbar\omega-(E_n-E_0)+i\eta}-
   {1\over\hbar\omega+(E_n-E_0)-i\eta}\right],
   \label{eq:piqq}
\end{eqnarray}
where
\begin{equation}
\hat H|\psi_n>=E_n|\psi_n>,
\end{equation}
$\hat H$ being the full nuclear Hamiltonian and ${\hat O}({\bf q})$
the second quantized expression of the vertex operator.
We shall mainly be concerned with the $\sigma\tau$ nuclear
excitations: in such a case one has
\begin{mathletters}
\begin{equation}
  O_L({\bf q},{\bf r})=\tau_\alpha
  \mbox{\boldmath $\sigma$}\cdot\hat{\bf q}
  {\rm e}^{i{\bf q}\cdot{\bf r}}
  \label{eq:ol}
\end{equation}
in the {\it longitudinal} channel and
\begin{equation}
  O_T({\bf q},{\bf r})={\tau_\alpha\over\sqrt{2}}
  \mbox{\boldmath $\sigma$}\times\hat{\bf q}
  {\rm e}^{i{\bf q}\cdot{\bf r}}
  \label{eq:ot}
\end{equation}
\end{mathletters}
in the {\it transverse} one.
In the following, for sake of convenience, we shall set $E_0=0$.

The angular part of $\Pi({\bf q},{\bf q}';\omega)$ can be
handled through a multipole decomposition which reads \cite{alb85}
\begin{equation}
  \Pi_L({\bf q},{\bf q}';\omega)=
  \sum_{JM}\Pi_J(q,q';\omega)
  Y^*_{JM}(\hat{\bf q})Y_{JM}(\hat{\bf q}')
  \label{eq:pil}
\end{equation}
in the {\boldmath $\sigma$}$\cdot{\bf q}$ channel and
\begin{equation}
  \Pi_T({\bf q},{\bf q}';\omega) =
  \sum_{JJ'M'} \Pi_{JJ'}(q,q';\omega)
  Y^*_{J'M'}(\hat{\bf q})Y_{J'M'}(\hat{\bf q}'){2J+1\over2J'+1}
  \label{eq:pit}
\end{equation}
in the {\boldmath $\sigma$}$\times{\bf q}$ one.

The latter expression, somewhat more involved than the former, is
shortly derived in Appendix~\ref{app:A}.
Also
\begin{mathletters}
\begin{equation}
  \Pi_J(q,q';\omega)=
  \sum_{\ell\ell'}\left[{\hat\Pi}_J(q,q';\omega)\right]_{\ell\ell'}
  a_{J\ell}a_{J\ell'}
\end{equation}
and
\begin{equation}
  \Pi_{JJ'}(q,q';\omega)={1\over2}\sum_{\ell\ell'}
     \left[{\hat\Pi}_J(q,q';\omega)\right]_{\ell\ell'}
     b^{J'}_{J\ell}\,b^{J'}_{J\ell'},
  \label{eq:pijj}
\end{equation}
\end{mathletters}
where
\begin{mathletters}
\begin{equation}
  a_{J\ell}=(-1)^\ell\sqrt{2\ell+1}\left(
  \begin{array}{ccc}
  \ell&1&J\\ 0&0&0
  \end{array}
  \right).
  \label{eq:ajl}
\end{equation}
and
\begin{equation}
  b^{J'}_{J\ell}=\sqrt{6}\sqrt{2J'+1}\,a_{J'\ell}
   \left\{
   \begin{array}{ccc}
    J&J'&1\\ 1&1&\ell
   \end{array}
   \right\}.
\end{equation}
\end{mathletters}

{}From (\ref{eq:pil}) and (\ref{eq:pit}) the following expressions
for the $\sigma\tau$ nuclear responses are then obtained:
\FL
\begin{eqnarray}
  R_{L,T}(q,\omega) &=& -{1\over\pi}{\rm Im}\Pi_{L,T}(q,q,;\omega)
  \nonumber\\
  &=& -{1\over4\pi^2}{\rm Im}\sum_J(2J+1)\Pi_{J(L,T)}(q,q;\omega),
  \label{eq:rlt}
\end{eqnarray}
with
\begin{mathletters}
\begin{eqnarray}
  \Pi_{J(L)}(q,q';\omega) &\equiv & \Pi_J(q,q';\omega)\nonumber\\
    &=& \sum_{\ell\ell'}
    \left[{\hat\Pi}_J(q,q';\omega)
    \right]_{\ell\ell'}a_{J\ell}a_{J\ell'}
\end{eqnarray}
and
\FL
\begin{eqnarray}
  \Pi_{J(T)}(q,q';\omega) &=& \sum_{J'}\Pi_{JJ'}(q,q';\omega)
  \nonumber\\
    &=& {1\over2}\sum_{\ell\ell'}
    \left[{\hat\Pi}_J(q,q';\omega)\right]_{\ell\ell'}
    (\delta_{\ell\ell'}-a_{J\ell}a_{J\ell'}).
  \label{eq:pijt}
\end{eqnarray}
\end{mathletters}
The evaluation of the $J$-th multipole ${\hat\Pi}_J$, the
{\it dynamical part} of the polarization propagator, is carried
out in this paper in the RPA framework by solving the following
set of coupled (by the tensor interaction) integral equations
\cite{alb85}
\widetext
\begin{eqnarray}
   \left[{\hat\Pi}_J^{\rm RPA}(q,q';\omega)\right]_{\ell\ell'} &=&
    \left[{\hat\Pi}_J^0(q,q';\omega)\right]_{\ell\ell'} \nonumber\\
   && +{1\over(2\pi)^3}\int_0^\infty dk\,k^2\sum_{\ell_1\ell_2}
    \left[{\hat\Pi}_J^0(q,k;\omega)\right]_{\ell\ell_1}
    \left[U_J(k,\omega)\right]_{\ell_1\ell_2}
    \left[{\hat\Pi}_J^{\rm RPA}(k,q';\omega)\right]_{\ell_2\ell'}.
   \nonumber\\
   \label{eq:pijrpa}
\end{eqnarray}
The bare ph polarization propagator (obtained by replacing the full
nuclear
hamiltonian $\hat H$ with the mean field one ${\hat H}^0$) reads
\begin{mathletters}
\begin{equation}
  \left[{\hat\Pi}_J^0(q,q';\omega)\right]_{\ell\ell'} =
  \sum_{ph} Q^{J\ell}_{ph}(q)
  \left[{1\over\hbar\omega-(\epsilon_p-\epsilon_h)+i\eta}
   -{1\over\hbar\omega+(\epsilon_p-\epsilon_h)-i\eta}\right]
   {Q^{J\ell'}_{ph}}^*(q'),
  \label{eq:pi0ll}
\end{equation}
where
\begin{eqnarray}
  Q^{J\ell}_{ph}(q) &=&
  < \!j_p j_h;J|\ell\sigma;J\!>\delta_{\sigma,1}(-i)^{\ell+1}
  (-1)^{\ell_h}4[\pi(2\ell_p+1)(2\ell_h+1)]^{1/2} \nonumber\\
  && \times {\cal I}_{\ell n_p\ell_p j_p n_h\ell_h j_h}(q)
  \left(
  \begin{array}{ccc}
  \ell_p&\ell_h&\ell\\ 0&0&0
  \end{array}
  \right).
  \label{eq:qjl}
\end{eqnarray}
\narrowtext
\noindent
In Eq.~(\ref{eq:qjl}) $<\! j_p j_h;J|\ell\sigma;J \!>$ is the
standard $LS-jj$
recoupling coefficient
and\footnote{With respect to previous work \cite{alb85},
  we have changed here the definition of $\hat O_{L,T}$ in
  (\ref{eq:ol}) and (\ref{eq:ot})
  by using the unit versor $\hat{\bf q}$ instead of the vector
  ${\bf q}$   in the spin longitudinal and transverse operators.
  Accordingly, the quantity ${\cal I}$ is now defined without a
  factor   $q$ in front whereas in the ph interaction (\ref{eq:vlt})
  appears an additional factor $k^2$.
  Also, a factor $1/\sqrt2$ has been introduced in   the
  (\ref{eq:ot}) in order to have the same normalization for both the
  longitudinal and transverse propagators. }
\FL
\begin{equation}
  {\cal I}_{\ell n_p\ell_p j_p n_h\ell_h j_h}(q)=\int_0^\infty
  dr\,r^2j_\ell(qr)
  R_{n_p\ell_p j_p}(r)R_{ n_h\ell_h j_h}(r).
  \label{eq:il}
\end{equation}
\end{mathletters}
If the spin-orbit term of the nuclear mean field is neglected, then
(\ref{eq:pi0ll}) is diagonal in the orbital angular momentum, i.~e.
$[{\hat\Pi}_J^0(q,q';\omega)]_{\ell\ell'}
  =\delta_{\ell\ell'}\hat\Pi^0_\ell(q,q';\omega)$, where
\widetext
\begin{eqnarray}
  \hat\Pi^0_\ell(q,q';\omega) &=&
  16\pi\sum_{\scriptstyle n_p\ell_p\atop
   n_h\ell_h} (2\ell_p+1)(2\ell_h+1)
  \left(\begin{array}{ccc}
  \ell_p&\ell_h&\ell\\ 0&0&0
  \end{array}\right)^2
  {\cal I}_{\ell n_p\ell_p n_h\ell_h}(q)\,
  {\cal I}_{\ell n_p\ell_p n_h\ell_h}(q')\nonumber\\
  &&\times
  \left[{1\over\hbar\omega-
    (\epsilon_{n_p\ell_p}-\epsilon_{ n_h\ell_h})+i\eta}-
    {1\over\hbar\omega+(\epsilon_{n_p\ell_p}-
    \epsilon_{ n_h\ell_h})-i\eta}\right].
\end{eqnarray}
\narrowtext
\noindent
In (\ref{eq:pi0ll}), (\ref{eq:qjl}) and (\ref{eq:il}) $R_{p(h)}$
are the radial wave functions and $\epsilon_{p(h)}$ the associated
eigenvalues. They are obtained by solving the Schroedinger
equation with the Woods-Saxon potential
\begin{equation}
  W(r) = {W_0\over1+{\rm e}^{(r-R)/a}}
   +\left[{\hbar c\over m_\pi^2 c^2}\right]^2{W_{so}\over a r}
   {{\rm e}^{(r-R)/a}\over\left[1+{\rm e}^{(r-R)/a}\right]^2}\,
   \mbox{\boldmath $\ell\cdot\sigma$}
   \label{eq:ws}
\end{equation}
(in the present paper, for sake of simplicity, the Coulomb
potential is neglected), where $m_\pi$ is the pion mass.
Obviously, for the states in the continuum part of the spectrum
the sum over $n_p$ has to be changed into an integral, since the
principal quantum number $n_p$ becomes a continuum variable.
Instead of following the standard procedure of calculating the
propagator in the coordinate space and then Fourier transform to
the momentum space, we have chosen to evaluate directly
Eq.~(\ref{eq:pi0ll}): the calculation is fast and straightforward
and one only needs to take some care in performing the
integrals over the ph energies in the resonance region.

As already mentioned, in the RPA equations we employ the ph
interaction $g'+V_\pi+V_\rho$, namely
\begin{equation}
  \left[U_J(k,\omega)\right]_{\ell_1\ell_2} =
  V_L(k,\omega)a_{J\ell_1}a_{J\ell_2}
  +V_T(k,\omega)(\delta_{\ell_1\ell_2}-a_{J\ell_1}a_{J\ell_2}),
\end{equation}
where
\FL
\begin{eqnarray}
  V_L(k,\omega) &=&
  \Gamma^2_\pi(k_\mu^2){f^2_\pi\over\mu_\pi^2}\left[g'+
   {k^2\over\omega^2-k^2-\mu^2_\pi}\right] \nonumber\\
   \label{eq:vlt}\\
  V_T(k,\omega) &=&
  \left[\Gamma^2_\pi(k_\mu^2){f^2_\pi\over\mu_\pi^2}g'+
   \Gamma^2_\rho(k_\mu^2){f^2_\rho\over\mu_\rho^2}
   {k^2\over\omega^2-k^2-\mu^2_\rho}\right].
   \nonumber
\end{eqnarray}
In the above, $f^2_\pi/4\pi\hbar c=0.08$, $f^2_\rho/\mu^2_\rho=
  2.18(f^2_\pi/\mu^2_\pi)$, $k_\mu\equiv(\omega,{\bf k})$
and the usual monopole form factors
\begin{equation}
  \Gamma_{\pi,\rho}(k_\mu^2)={\Lambda_{\pi,\rho}^2-m_{\pi,\rho}^2
  \over\Lambda_{\pi,\rho}^2-\omega^2+k^2}
\end{equation}
have been included at the $\pi$($\rho$)NN vertices with cut-offs
$\Lambda_\pi=1300$~MeV and $\Lambda_\rho=1700$~MeV, respectively.
Moreover, in the following we shall always use $g'=0.7$.

Since we shall also be interested in (p,p$'$) reactions, we have
to introduce, in addition to the spin-isospin responses, a response
function $R_{ST}$ in the scalar ($S=0$) and isoscalar ($T=0$)
channel ((p,p$'$) scattering at intermediate energies is in fact
largely dominated by this channel). This response reads
\begin{equation}
$$ R_{00}(q,\omega)=-{1\over4\pi^2}\mbox{\rm Im}\sum_{J}(2J+1)
  \Pi_{J(00)}(q,q;\omega),
\end{equation}
where at zero-order $\Pi_{J(00)}(q,q';\omega)$ coincides with the
$[\hat\Pi^0_{J}(q,q';\omega)]_{JJ}$ of Eq.~(\ref{eq:pi0ll}),
substituting
$\delta_{\sigma,1}$ with $\delta_{\sigma,0}$ in (\ref{eq:qjl}).
In RPA also $\Pi_{J(00)}$ obeys an integral equation similar to
(\ref{eq:pijrpa}) and for the ph interaction we have used the
$G$-matrix of Ref.~\cite{muth87} at a density of roughly one-half
the central nuclear density.

\subsection{The Spreading Width}
\label{subsec:spread}

The continuum RPA framework naturally accounts for the
{\it escape width} of the particle states.
However, also the {\it spreading width} of the
latter plays an important role in the nuclear many-body problem and
should be reckoned with. For this purpose, it is convenient,
following Ref.~\cite{wambach88}, to recast the expression
(\ref{eq:piqq}) for the polarization propagator as follows
\begin{equation}
  \Pi({\bf q},{\bf q}';\omega)=
  <\psi_0|{\hat O}({\bf q})\,
  G(\omega)\,{\hat O}^\dagger({\bf q}')|\psi_0>,
\end{equation}
where
\begin{equation}
  G(\omega)=
  {1\over\hbar\omega-\hat H+i\eta}-
  {1\over\hbar\omega+\hat H-i\eta}
\end{equation}
is the propagator of the excitation induced on the exact nuclear
ground state $|\psi_0>$ by the operator ${\hat O}({\bf q})$.

Now, in the framework of Feshbach's formalism, starting from
\begin{equation}
  \hat H=\hat H_0+\hat V
\end{equation}
one can derive the following effective Hamiltonian, restricted to
operate in the ph space:
\begin{eqnarray}
  \hat H_{\it eff} &=& P\hat HP-P\hat VQ{1\over Q\hat
   HQ-\hbar\omega-i\eta}Q\hat VP \nonumber\\
  &=& P\hat HP-U(\omega) \nonumber\\
  &=& \hat H^0_{\it eff}+P\hat VP,
\end{eqnarray}
having set
\begin{equation}
  \hat H^0_{\it eff}=P\hat H_0P-U(\omega).
\end{equation}
In the above $P$ and $Q$ are the projection operators in the ph
and in the complement space, respectively.

Therefore, to allow the formalism to encompass the coupling of the
ph excitations to the 2p2h ones (or to more complicated
configurations), i.~e. to incorporate the spreading width of the ph
states, one should replace in the previously outlined RPA scheme the
``bare'' ph propagator with the expression
\FL
\begin{eqnarray}
  \Pi^0({\bf q},{\bf q}';\omega) &=&
   \sum_{ph}  <\phi_0|{\hat O}({\bf q})|\phi_{ph}>
   <\phi_{ph}|{\hat O}^\dagger({\bf q}')|\phi_0>\nonumber\\
  && \times\left[{1\over\hbar\omega+\Sigma_{ph}(\omega)-
   (\epsilon_p-\epsilon_h)+i\eta}
   -{1\over\hbar\omega-\Sigma_{ph}(-\omega)+(\epsilon_p-\epsilon_h)
   -i\eta}\right],\nonumber\\
\end{eqnarray}
\noindent
where
\begin{equation}
  \Sigma_{ph}(\omega)=
  <\phi_{ph}|\hat VQ(Q\hat HQ-\hbar\omega-i\eta)^{-1}Q\hat V
  |\phi_{ph}>
\end{equation}
and
\begin{equation}
  \hat H_0|\phi_{ph}>=(\epsilon_p-\epsilon_h)|\phi_{ph}>,
\end{equation}
the $|\phi_{ph}>$ obviously being the ``bare'' ph states.

A microscopic calculation of $\Sigma_{ph}(\omega)$ is difficult
to carry out. An alternative approach \cite{wambach88}, to which
we shall adhere, assumes that the real and the imaginary parts of
$\Sigma_{ph}$, namely
\begin{equation}
  \Sigma_{ph}(\omega)=
  \Delta_{ph}(\omega)+i{\Gamma_{ph}(\omega)\over2},
\end{equation}
can be cast into the form
\begin{eqnarray}
  \Gamma_{ph}(\omega) &=&
  \gamma_p(\hbar\omega+\epsilon_h)+\gamma_h(\epsilon_p-\hbar\omega)
  \nonumber\\
  \\
  \Delta_{ph}(\omega)
  &=&
  \Delta_p(\hbar\omega+\epsilon_h)+\Delta_h(\epsilon_p-\hbar\omega),
  \nonumber
\end{eqnarray}
where the arguments of the functions appearing in the right hand
side of the above expressions are inferred from the analysis of
the second order diagrams of Fig.~\ref{fig1}(a).
Note that the diagrams corresponding to the exchange of a bubble
between two fermionic lines, conjectured to be small in the
$\sigma\tau$ channel, are neglected (Fig.~\ref{fig1}(b)).

Now, instead of calculating these diagrams, we utilize the formulae
\begin{eqnarray}
  \gamma_p(\epsilon) &=&
   2\alpha\left({\epsilon^2\over\epsilon^2+\epsilon^2_0}\right)
   \left({\epsilon^2_1\over\epsilon^2+\epsilon^2_1}\right)
   \theta(\epsilon)\nonumber\\
  \label{eq:gamph} \\
   \gamma_h(\epsilon) &=& \gamma_p(-\epsilon),\nonumber
\end{eqnarray}
symmetrical with respect to the Fermi energy ($\epsilon_F=0$),
which give a reasonable fit of the particle widths for
medium-heavy nuclei, using $\alpha=10.75$ MeV, $\epsilon_0=18$ MeV
and $\epsilon_1=110$ MeV \cite{mahaux81}.
We then insert these expressions into the once-subtracted
dispersion relation
\begin{equation}
  \Delta_p(\epsilon)={\epsilon\over2\pi}\,{\rm P} \int_0^\infty
  d\epsilon'\,
  {\gamma_p(\epsilon')\over(\epsilon'-\epsilon)\epsilon'}\,,
\end{equation}
(and in a similar one for $\Delta_h$) obtaining
\begin{eqnarray}
  \Delta_p(\epsilon) &=&
  {1\over\pi}{\alpha\epsilon_1^2\over\epsilon_1^2
    -\epsilon_0^2}
  \left\{{\epsilon\over2}
  \left[{\epsilon_0\over\epsilon^2+\epsilon_0^2}
       -{\epsilon_1\over\epsilon^2+\epsilon_1^2}\right]
   -\epsilon^2\left[
    {\log|\epsilon/\epsilon_0|\over\epsilon^2+\epsilon_0^2}
    -{\log|\epsilon/\epsilon_1|
     \over\epsilon^2+\epsilon_1^2}\right]\right\}
   \nonumber\\
  \label{eq:delph}\\
   \Delta_h(\epsilon)&=&-\Delta_p(-\epsilon).\nonumber
\end{eqnarray}
The subtraction at the Fermi surface avoids the double
counting of the smooth background in the single particle energy
already embodied in the mean field.

Formulae (\ref{eq:gamph}) and (\ref{eq:delph}) are the ones we shall
employ in analyzing the $\sigma\tau$ nuclear response functions,
dressing the bare ph propagator and then solving the RPA equations.

\section{Surface Response Functions}
\label{sec:surf}

\subsection{The Response Function in the One-Step Glauber Theory}
\label{subsec:onestep}

In order to treat the response of the nucleus to a strongly
interacting probe, one should enlarge the framework described in
the previous Section to account for both the distortion and the
absorption of the impinging probe on the surface of the nucleus.

In the framework of Glauber theory \cite{glauber59}, the
scattering amplitude of a probe on a nucleus of mass number $A$
is given by
\begin{equation}
  F_{fi}({\bf q})=
  {ik\over2\pi}\int d{\bf b}\, e^{i{\bf q}\cdot{\bf b}}
  <\!\psi_f|\Gamma({\bf b};{\bf s}_1...{\bf s}_A)|\psi_i\!>,
  \label{eq:ffi}
\end{equation}
where the so-called nuclear profile function $\Gamma$ is
expressed as
\begin{equation}
  \Gamma({\bf b};{\bf s}_1...{\bf s}_A)=
  1-\prod_{j=1}^A[1-\Gamma_j({\bf b}-{\bf s}_j)]
\end{equation}
in terms of the single nucleon profile function
\begin{equation}
  \Gamma_j({\bf b})={1\over2\pi ik}\int d
  \mbox{\boldmath $\lambda$}
   \,e^{-i\mbox{\boldmath $\scriptstyle \lambda$}\cdot{\bf b}}
   f_j(\mbox{\boldmath $\lambda$}).
  \label{eq:gamj}
\end{equation}
In the above, ${\bf b}$ is the impact parameter and
${\bf q}={\bf k}-{\bf k}'$ the transferred momentum (${\bf k}$
and ${\bf k}'$ are the projectile incoming and outgoing momenta,
respectively): they are bi-dimensional vectors in the
plane orthogonal to the direction of motion of the projectile.
The probe-nucleon amplitudes $f(\mbox{\boldmath $\lambda$})$
of Eq.~(\ref{eq:gamj}) are assumed to be the free ones and are
meant to be evaluated in the laboratory system.

If the excitation energy of the nucleus is supplied by the probe
in a single collision (one-step assumption), then, for large
enough nuclei, one can rewrite Eq.~(\ref{eq:ffi}) in the form
\FL
\begin{equation}
  F_{fi}({\bf q})={ik\over2\pi}\int d{\bf b}\,
  e^{i{\bf q}\cdot{\bf b}}
  e^{i\chi_{\rm opt}(b)}
  <\!\psi_f|\Gamma_{\sigma\tau}({\bf b}-{\bf s})|\psi_i\!>,
  \label{eq:ffii}
\end{equation}
where, for definiteness, we consider the spin-isospin inelastic
channel. The complex phase shift $\chi_{\rm opt}$, responsible
for the absorption and distortion of the probe, is given by
\begin{equation}
  \chi_{\rm opt}(b)={2\pi\over k}f(0)\,T(b),
\end{equation}
with
\begin{equation}
  T(b)=\int_{-\infty}^{+\infty}  dz\,\rho(r=\sqrt{b^2+z^2}),
\end{equation}
$\rho(r)$ being the nuclear density (assumed in the following to
be represented by a Fermi distribution) and $f(0)$ the forward
total NN scattering amplitude. At high energies, the imaginary
part of $f(0)$ is dominant and, therefore, making use of the
optical theorem, one can write
\begin{equation}
  \chi_{\rm opt}(b)={i\over 2}\widetilde{\sigma}_{\rm tot}\,T(b),
  \label{eq:chiopt}
\end{equation}
where $\widetilde{\sigma}_{\rm tot}$ is the effective
probe-nucleon total cross-section (effective because empirically
embodying Pauli blocking effects).

Eq.~(\ref{eq:ffii}) suggests to replace the vertex operators
$O_{L,T}({\bf q},{\bf r})$ of (\ref{eq:ol}) and (\ref{eq:ot})
with the new ones
\FL
\begin{equation}
  O^{\it surf}_{L,T}({\bf q},{\bf r}) = {1\over(2\pi)^2 f_{L,T}(q)}
  \int d{\bf b}\,d \mbox{\boldmath $\lambda$}\,
  e^{i\chi_{\rm opt}(b)}
  e^{i({\bf q}-\mbox{\boldmath $\scriptstyle \lambda$})\cdot{\bf b}}
  f_{L,T}(\lambda) O_{L,T}(\mbox{\boldmath $\lambda$},{\bf r}),
  \label{eq:olts}
\end{equation}
where the $f_{L,T}(\lambda)$ are the elementary isovector
spin-longitudinal and spin-transverse probe nucleon scattering
amplitudes. From the above expression one sees that the probe does
not transfer a single momentum ${\bf q}$ to the nucleus, but rather
all possible momenta {\boldmath$\lambda$} with weight
$f_{L,T}(\lambda)$. Furthermore, its distortion is controlled by
the factor $\exp[i\chi_{\rm opt}(b)]$.
The normalization in Eq.~(\ref{eq:olts}) has been chosen in order
to recover the standard vertex operators in the limit of no
distortion:
\begin{equation}
  O^{\it surf}_{L,T}\:
  \stackrel{\tilde\sigma_{\rm tot}\rightarrow0}{\longrightarrow}\:
  O_{L,T}.
\end{equation}

For sake of simplicity, we have not explicitly expressed the
dependence of $O^{\it surf}_{L,T}$ on the spin-isospin operators
of the probe (see, however, Appendix~\ref{app:B}).
If we insert the pertinent surface vertex operators in
Eq.~(\ref{eq:piqq}) and take the appropriate matrix elements of the
spin and isospin probe operators, we can finally define two new
polarization propagators
$\Pi^{\it surf}_{L,T}({\bf q},{\bf q}';\omega)$.
The associated one-step response functions,
$R^{(1)\it surf}_{L,T}(q,\omega)$, can then be obtained by
substituting, in Eq.~(\ref{eq:rlt}), $\Pi_{J(L,T)}(q,q;\omega)$
with the corresponding surface expressions,
$\Pi^{\it surf}_{J(L,T)}(q,q;\omega)$, given by
\widetext
\begin{mathletters}
\begin{eqnarray}
  \Pi^{\it surf}_{J(L)}(q,q;\omega) &=& \Pi_{J(L)}(q,q;\omega)
  \nonumber\\
  && +{1\over|f_L(q)|^2}\int_0^\infty
  d\lambda\,\lambda\int_0^\infty
  d\lambda'\,\lambda'\,{\rm Re}[f^*_L(\lambda)\,
  f^{\phantom{*}}_L(\lambda')\,
  G_J(\lambda,\lambda';q)] \Pi_{J(L)}(\lambda,\lambda';\omega)
  \nonumber\\
  && -2{1\over|f_L(q)|^2}\int_0^\infty
  d\lambda\,\lambda\,{\rm Re}[f^*_L(q)\,
  f^{\phantom{*}}_L(\lambda)\,
  H_J(\lambda;q)]\Pi_{J(L)}(q,\lambda;\omega)
  \label{eq:pijls}
\end{eqnarray}
and
\begin{eqnarray}
  \Pi^{\it surf}_{J(T)}(q,q;\omega) &=& \Pi_{J(T)}(q,q;\omega)
  \nonumber\\
  && +{1\over|f_T(q)|^2}\int_0^\infty
  d\lambda\,\lambda\int_0^\infty
  d\lambda'\,\lambda'\,\sum_{J'}
  {\rm Re}[f^*_T(\lambda)\,f^{\phantom{*}}_T(\lambda')\,
  G_{J'}(\lambda,\lambda';q)] \Pi_{JJ'}(\lambda,\lambda';\omega)
  \nonumber\\
  && -2{1\over|f_T(q)|^2}\int_0^\infty
  d\lambda\,\lambda\,\sum_{J'}{\rm Re}[
  f^*_T(q)\,f^{\phantom{*}}_T(\lambda)\,H_{J'}(\lambda;q)]
  \Pi_{JJ'}(q,\lambda;\omega).
  \label{eq:pijts}
\end{eqnarray}
\end{mathletters}
\narrowtext
\noindent
In Eqs.~(\ref{eq:pijls}) and (\ref{eq:pijts}) we have set
\begin{eqnarray}
  G_J(\lambda,\lambda';q) &=& \sum_{\ell m} c_{J \ell m}
  g^*_m(\lambda,q) g_m(\lambda',q) \nonumber\\
  \\
   H_J(\lambda;q) &=& \sum_{\ell m}
   c_{J \ell m}g_m(\lambda,q),\nonumber
\end{eqnarray}
where
\FL
\begin{equation}
  g_m(\lambda,q)=\int_0^\infty  db\,b
  \left\{1-\exp[i\chi_{\rm opt}(b)]
  \right\}J_m(\lambda b)\,J_m(qb)\
  \label{eq:gm}
\end{equation}
and
\begin{eqnarray}
  c_{J \ell m} &=& I_{\ell+m}a_{J \ell}^2
  {(\ell-m-1)!!(\ell+m-1)!!\over(\ell+m)!!(\ell-m)!!}
  \nonumber\\
  \label{eq:cjl}\\
  I_{\ell+m} &=& \left\{
     \begin{array}{ccc}
     0,& \ell+m &\mbox{\rm odd} \\ 1,& \ell+m &\mbox{\rm even}
     \end{array}
     \right.\:.
  \nonumber
\end{eqnarray}
$\Pi_{JJ'}$ has been defined in Eq.~(\ref{eq:pijj}) or
(\ref{eq:pijt}). In Appendix~\ref{app:B}, we shortly sketch the
derivation of Eqs.~(\ref{eq:pijls}) and (\ref{eq:pijts}).

Again, also in the scalar-isoscalar channel, one can introduce
a surface propagator $\Pi^{\it surf}_{J(00)}$: it is easily
verified that its expression is identical to Eq.~(\ref{eq:pijls}),
replacing everywhere $(L)$ with $(00)$ ($f_{00}$ is the $S=0$,
$T=0$ NN amplitude) and setting
$a^2_{J\ell}\rightarrow\delta_{J,\ell}$ in (\ref{eq:cjl}).

\subsection{The Response Function in the Two-Step Glauber Theory}
\label{subsec:twostep}

Multiple scattering contributions to the nuclear response function
can be calculated along lines similar to the one-step case.
However, the problem gets numerically rather involved and therefore
we resort to an approximation that is often employed in Glauber
calculations.

In this approximation, each contribution from the multiple
scattering series to the response function is expressed as a volume
response function times a distortion factor
\cite{bertsch82,wambach87}. The latter is independent of the
transferred energy and momentum and should embody the distortion
effects. Thus, for instance, the one-step response can be written as
\begin{equation}
  R^{(1)}_{L,T}(q,\omega)={\cal D}_1\, R_{L,T}(q,\omega),
  \label{eq:r1lt}
\end{equation}
where $R_{L,T}(q,\omega)$ is the volume response of
Eq.~(\ref{eq:rlt}) and
\begin{equation}
  {\cal D}_1=N^{(1)}_{\rm eff}=\int  d{\bf b}\,T(b)
  e^{-\tilde{\sigma}_{\rm tot} T(b)},
  \label{eq:d1}
\end{equation}
$N^{(1)}_{\rm eff}$ being the effective number of nucleons
participating in the single collision.

A word of caution is in order here: expression (\ref{eq:r1lt}) is
able to reproduce the gross features of the nuclear response
function (say, its size), but yields the same shape of the volume
response.
This is at variance, as we shall see in the next Section
(see also Ref.~\cite{adp91}), with the full Glauber calculation of
the previous Subsection, since there the response is also reshaped,
a feature that cannot be overlooked, if one has to disentangle
genuine nuclear correlations from distortion effects.
This may be also the case, of course, when one comes to multistep
processes. Anyway, the size of their contribution, although not
negligible, is considerably smaller than the one arising from the
one-step scattering, thus reducing the sensitivity of the response
to the details of their shape.
This assumption has, however, to be assessed (see next Section).

For charge-exchange reactions at intermediate energies the two-step
response function can then be defined as \cite{wambach87}
\FL
\begin{eqnarray}
  R^{(2)}_{L,T}(q,\omega) &=&
  {{\cal D}_2\over k^2}{2\over|f_{L,T}(q)|^2}
    \int d{\bf q}' \int_0^\omega d\omega'
  \big|f_{L,T}(q')\big|^2 R_{L,T}(q',\omega') \nonumber\\
  && \times\big|f_{00}(|{\bf q}-{\bf q}'|)\big|^2
     R_{00}(|{\bf q}-{\bf q}'|,\omega-\omega'),
  \label{eq:r2lt}
\end{eqnarray}
\noindent
where ${\bf q}'$ is again a bi-dimensional vector, $k$ is the
momentum of the projectile and
\begin{equation}
  {\cal D}_2={1\over2}\int  d{\bf b}\,T^2(b)
  e^{-\tilde{\sigma}_{\rm tot}T(b)}
\end{equation}
is connected to the effective number of pairs participating in
the double scattering according to
\begin{equation}
  N^{(2)}_{\rm eff}=
  \left(
  \begin{array}{c}
  A\\ 2
  \end{array}
  \right)
  {\cal D}_2(\widetilde{\sigma}_{\rm tot})
    \big/{\cal D}_2(0).
\end{equation}
In Eq.~(\ref{eq:r2lt}), the charge-exchange reaction, driven by the
spin-isospin amplitudes, can occur only once and the second
scattering is driven by the scalar-isoscalar amplitudes.
The factor 2 comes from the two possible orderings of the reaction.

Non-charge-exchange reactions, on the other hand, are dominated by
the scalar-isoscalar channel, leading to the following definition:
\FL
\begin{eqnarray}
  R^{(2)}_{00}(q,\omega) &=&
  {{\cal D}_2\over k^2}{1\over|f_{00}(q)|^2}
    \int d{\bf q}' \int_0^\omega d\omega'
    \big|f_{00}(q')\big|^2 R_{00}(q',\omega') \nonumber\\
  && \times\big|f_{00}(|{\bf q}-{\bf q}'|)\big|^2
    R_{00}(|{\bf q}-{\bf q}'|,\omega-\omega').
  \label{eq:r200}
\end{eqnarray}
\noindent
Note that by inspecting (\ref{eq:r2lt}) and (\ref{eq:r200}) one
can already predict that two-step contributions will be twice more
important in charge-exchange reactions than in non-charge-exchange
ones. Indeed, the fact that in $R^{(2)}_{L,T}$ one of the
two rescatterings is driven by $|f_{00}|^2$ makes $R^{(2)}_{L,T}$
and $R^{(2)}_{00}$ practically of the same size, apart from the
factor 2 due to the two orderings of the charge-exchange reaction.

\section{Results}
\label{sec:results}

Let us start by discussing the {\it volume} responses, relevant to
electron scattering. We have performed the calculations for
$^{12}$C and $^{40}$Ca, using for the Woods-Saxon potential
(\ref{eq:ws}) the following set of parameters:
\begin{equation}
\begin{array}{rclrcl}
  W_0 &=& -54.8\,\mbox{\rm MeV}, & \: W_{so}
  &=& -10\,\mbox{\rm MeV},\\
  R   &=& 1.27\,A^{1/2}\,\mbox{\rm fm}, &\: a
  &=& 0.67\,\mbox{\rm fm}.
\end{array}
\end{equation}
In Fig.~\ref{fig2} we display the $\sigma\tau$ longitudinal and
transverse response functions of $^{40}$Ca at two values of the
momentum transfer $q$, comparing the free and RPA responses with
and without inclusion of the spreading width of the ph states.
When the spreading width is included, one observes a sizable
damping of the QEP and of the low-energy resonances, together
with a broadening of their width (since the energy sum rule has
to be conserved). The effect is essentially the same in the free
as in the RPA responses.

Note, as already mentioned in the Introduction, that in spite of
the different approximations for the inclusion of the collisional
damping of the single-particle motion underlying our approach and
the one of Ref.~\cite{wambach88}, the two calculations are in fact
in substantial agreement. Note also that these correlations,
related to the single-particle motion, affect the position of the
QEP much less than the RPA ones.

The transverse response function enters directly into the
expression for the nuclear transverse structure function, measured
in electron scattering experiments. Indeed, one has
\begin{equation}
  S_T(q,\omega) =
  {\mu_0^2\over e^2}(\mu_p-\mu_n)^2G_M^2(q_\mu^2)R_T(q,\omega),
  \label{eq:st}
\end{equation}
$\mu_0$ being the nuclear Bohr magneton, $\mu_p=2.79$,
$\mu_n=-1.91$ and
\begin{equation}
  G_M(q_\mu^2) = {1\over[1+({\bf q}^2-\omega^2/c^2)/18.1\,
  \mbox{\rm fm}^{-2}]^2}
\end{equation}
the usual electromagnetic $\gamma$NN form factor.
In (\ref{eq:st}) the small isoscalar contribution has been
neglected.

In Fig.~\ref{fig3} we compare the calculations (with inclusion
of the spreading width) of the previous figure to the experimental
data \cite{meziani85}. It is quite clear that the RPA correlations
are successful in bringing the QEP position to the right place, but
they miss the correct $q$-dependence of the strength. It is unlikely
that this shortcoming be due to the specific model we employ for
the ph interaction: in our treatment the Landau-Migdal parameter
$g'$ incorporates the effect of the exchange diagrams in the RPA
series and this approximation is known to work well in the
spin-isospin channel \cite{arima89}.

Thus, to improve the accord with experiment, further contributions
to the response function should likely be looked for beyond the ph
Hilbert space of RPA, such as 2p-2h and meson exchange current
terms.
This is further suggested by the fact that in the high energy region
of the response some strength is missing even at high momenta,
where the strength of the QEP is fairly close to the data.
See, for instance, Ref.~\cite{alb91} for a treatment of 2p-2h
contributions in nuclear matter, where, indeed, a satisfactory
description of (e,e$'$) data is found.

Let us now turn to the main topic of this work, namely to the
reactions involving strongly interacting probes.
In this case, a new
feature of the response functions is related to the strong
distortion of the projectile, whose strength is set by the
effective total probe-nucleon cross-section of (\ref{eq:chiopt}).

In Figs.~\ref{fig4} and \ref{fig5} one can see a comparison of the
one-step longitudinal and transverse {\it volume} (no distortion)
and {\it surface} (distorted) responses for $^{12}$C and $^{40}$Ca
at two transferred momenta, $q=1.54$ fm$^{-1}$ and
$q=2.31$ fm$^{-1}$.
In each plot besides the volume response
($\widetilde{\sigma}_{\rm tot}=0$), two surface cases
($\widetilde{\sigma}_{\rm tot}=30$ mb and
$\widetilde{\sigma}_{\rm tot}=40$ mb) are displayed, for both the
free and the RPA responses.
As it may be expected, one observes a strong quenching of the
surface responses, together with a reduction of the importance of
the RPA correlations, since the reaction is now mainly confined at
the low density, peripheral region of the nucleus.

Furthermore, another effect shows up, of great importance for
the interpretation of the experimental data, namely a shift of
the QEP position of the surface responses with respect
to the volume ones: it appears to be sizable (due to the fact
that RPA correlations are damped, it is much larger than the shift
induced by the latter) and practically independent of
$\widetilde{\sigma}_{\rm tot}$ in any realistic range of values
for this parameter. It also turns out to be independent of the
elementary NN amplitudes (here we employ the parameterization of
Bugg and Wilkin \cite{bugg85}).

However, comparing the results for $^{12}$C and $^{40}$Ca, one
also observes a shell dependence for this effect: this is better
illustrated in Fig.~\ref{fig6}, where the surface longitudinal
response functions for $^{12}$C and $^{40}$Ca are directly
compared for the case $\widetilde{\sigma}_{\rm tot}=40$ mb at
$q=1.54$ fm$^{-1}$ and $q=2.31$ fm$^{-1}$, both with and without
inclusion of the spreading width.
In order to bring the responses at the same scale, $R_L$ has
been divided by $N_{\rm eff}$ (see (\ref{eq:d1})), with
$N_{\rm eff}(^{12}$C$)=3.2$ and $N_{\rm eff}(^{12}$Ca$)=5.6\;$,
respectively.

In Ref.~\cite{adp91} it has been shown that the $^{40}$Ca
one-step surface responses are hardened for $q\alt 1.8$ fm$^{-1}$
and softened for $q\agt 1.8$ fm$^{-1}$;
in contrast, $^{12}$C surface responses are always hardened, as
one can see from Fig.~\ref{fig6}. Furthermore, this feature is
independent of the specific value of $\widetilde{\sigma}_{\rm tot}$,
which is dictated by the energy and the nature of the projectile.

A contribution that might, in principle, reshape the quasi-elastic
response functions is the two-step term in the Glauber multiple
scattering expansion. In Ref.~\cite{wambach87} it had been shown
to produce at small transferred momenta a contribution smoothly
increasing with the transferred energy. However, this feature,
because of the NN amplitudes entering into (\ref{eq:r2lt}) and
(\ref{eq:r200}), is strongly dependent upon the momentum regime.

To figure it out, we plot in Fig.~\ref{fig7} the two-step
response $R^{(2)}$ for a simple model, namely assuming the two
squared amplitudes in (\ref{eq:r200}) to be equal and with a
gaussian shape, $|f(q)|^2=A\exp{(-\eta q^2)}$.
Here, and also in the following calculations with realistic NN
amplitudes, in order to estimate $R^{(2)}$ we use for simplicity
the free harmonic oscillator model without inclusion of the
spreading width: since $R^{(2)}$ is the convolution of two ph
response functions, it should be rather independent of the
details of their shape.

The curves in Fig.~\ref{fig7} correspond to $\eta$ ranging from
0.001 fm$^2$ to 1 fm$^2$ and it is quite apparent that the shape
changes when the $q$-dependence of the amplitudes becomes more
pronounced. The reason for this behaviour lies in (\ref{eq:r200}):
indeed, when $f(q')$ is a rapidly decreasing function, the main
contribution to the integrals comes from the region $q'\sim0$ and,
as a consequence, $\omega'\sim0$; it then follows that at fixed
$q$ the maximum of $R^{(2)}(q,\omega)$ will be found for $\omega$
around the QEP position.
Note that for $\eta\agt 0.1$ fm$^2$ one gets, at high momenta,
a maximum {\it below} the QEP and that realistic values for $\eta$,
which fit the NN amplitudes, fall just in this range.

Let us now see how our full model (namely one-step RPA Glauber
responses with spreading width plus two-step free responses)
compares to the available ($p$,$n$) data \cite{tadd91}.
The double differential cross-section for a ($p$,$n$) reaction
is given by
\begin{equation}
  {d^2\sigma\over d\Omega d\omega} =
  \sum_{\alpha=L,T} |f_\alpha(q)|^2
  \big[R^{(1)\it surf}_\alpha(q,\omega) +
  R^{(2)}_\alpha(q,\omega)\big]
  \label{eq:csa}
\end{equation}
and it is tested in Fig.~\ref{fig8} on the data from a reaction
on $^{12}$C at a proton energy of 795 MeV and for four different
scattering angles ($\theta=9^o$, $12^o$, $15^o$ and $18^o$),
corresponding to $q$ ranging from 1.16 fm$^{-1}$ to 2.31 fm$^{-1}$.

The dashed and the solid curves
in each plot represent the one-step and the full calculations,
respectively, whereas the dotted curve is the two-step term; for
sake of comparison, we display also the cross-section corresponding
to the simpler model based on the
one-step contribution of Eq.~(\ref{eq:r1lt}) (dot-dashed line),
i.~e.
\begin{equation}
  {d^2\sigma\over d\Omega d\omega} =
  \sum_{\alpha=L,T} |f_\alpha(q)|^2
  \big[R^{(1)}_\alpha(q,\omega) + R^{(2)}_\alpha(q,\omega)\big].
  \label{eq:csb}
\end{equation}

A few comments are in order. First, the reduction in strength in
going from (\ref{eq:csa}) to (\ref{eq:csb}) is due to RPA
correlations (indeed, the difference fades away at large $q$),
whereas the shift in the QEP position comes from the cylindrical
geometry of the reaction, which is respected by our approach
(it stays constant with $q$).
Looking at the data, it is apparent that now the situation is
complementary to what we have observed in the case of the
($e$,$e'$) data: in fact, we are now able to reproduce the correct
$q$-dependence of the strength, but not the QEP position of the
(p,n) data, which is still somewhat more hardened than predicted
by our calculations.

Note that the height of the peak depends on
$\widetilde\sigma_{\rm tot}$: here, we have used
$\widetilde\sigma_{\rm tot}=40$~mb, without accounting for
medium effects on this parameter.
The estimate of this effective parameter is not well assessed, since
there are large discrepancies between the direct calculation of
Pauli blocking effects \cite{bertsch82}\ and the derivation of
$\widetilde\sigma_{\rm tot}$ from the optical potential
\cite{wambach87}. Using $\widetilde\sigma_{\rm tot}=30$~mb, the
curves in Fig.~\ref{fig8} would get a 25\%\ increasing at any
momentum transfer, without affecting the $q$-dependence of the
strength and reproducing correctly also the height of the peak.

The real problem with the model we are discussing is related to the
$A$-dependence of the QEP position that it introduces
(see Fig.~\ref{fig6}), since the (p,n) data of Ref.~\cite{tadd91}
appear to scale, at least in $^{12}$C and Pb.
A scale factor is also the only difference in the ($^3$He,t)
cross-sections on $^{12}$C and $^{40}$Ca \cite{berg87}, whose QEP
position displays a pattern, as a function of $q$, quite different from
the (p,n) case, in agreement with our calculations for
$^{40}$Ca \cite{adp91}, but not for $^{12}$C. In this connection,
we remind the reader that in Ref.~\cite{adp91} the $^3$He projectile
has been treated as structureless, whereas, as already
mentioned, a proper account of its internal structure might affect
the shape of the response functions \cite{dmitr92}.

It is not clear, at the moment, whether this shell dependence of the
Glauber quasi-elastic responses is a genuine effect or, rather, it
reflects a shortcoming of the model or the approximations of our
treatment: indeed, in the two-step contribution we have neglected
the cylindrical geometry of the reaction, which, as we noted, is
the source of the shift in the one-step term.
Actually, the discrepancy between $^{12}$C and $^{40}$Ca shows up at
large momenta (see Fig.~\ref{fig6}), where the two-step response
is relatively more important.

Finally, in Fig.~\ref{fig9} we show the same comparison as in
Fig.~\ref{fig8}, but for (p,p$'$) scattering at 795 MeV
\cite{chrien80}.
We have considered only the dominant scalar-isoscalar
channel, which means that in (\ref{eq:csa}) and (\ref{eq:csb}) we
have set $\alpha=00$.
By inspecting Fig.~\ref{fig9} a few observations follow:
the two-step contribution is smaller than in the (p,n) case, as
anticipated in Sec.~\ref{subsec:twostep};
the simple calculation based on (\ref{eq:r1lt}) of course yields
a better account of the QEP position, since (p,p$'$) data do not
show any shift; on the other hand, the $q$-dependence of the
height of the peak is unsatisfactorily described, but this is due
to our choice of only allowing the scalar-isoscalar channel: when
the momentum transfer increases (say, over 1.5 fm$^{-1}$) other
components of the NN amplitudes become important and
should help in increasing the height of the response.

\section{Conclusions}
\label{sec:concl}

The aim of this paper has been to investigate the quasi-elastic
nuclear response as seen in electron and, especially, proton
inclusive scattering.
We restricted ourselves to unpolarized scattering, since
already in this domain a number of issues wait to be clarified.

Our attention has been mainly directed to the (p,n) and (p,p$'$)
reactions, owing to the surprising and apparently contradictory
figures displayed by these experiments.
We have first applied our formalism, based on continuum RPA plus
spreading width of the ph states, to the transverse electron
scattering finding that:
\begin{itemize}
\item[i)] RPA correlations shift the QEP position to higher energy;
\item[ii)] the $q$-dependence of the strength in the QEP domain is
not correctly reproduced by our model, since some strength is
actually missing at low but not at high momenta;
\item[iii)] in the high energy tail of the response strength is
missing at all momenta.
\end{itemize}

The solution of the problem in items ii) and iii) likely comes from
the contribution from 2p-2h processes and meson-exchange
currents, as can be seen from the results of Ref.~\cite{alb91},
worked out in nuclear matter.
Neglecting 2p-2h contributions in hadron scattering should be
a less serious shortcoming, because of the strong density
dependence, which makes them relatively less important in the low
density surface regions probed by strongly interacting projectiles.

In hadron scattering, on the other hand, multiple-scattering
contributions can be substantial and, accordingly, we have adopted
the framework of Glauber theory: one-step processes have been
calculated consistently within the theory, whereas for two-step
processes we have resorted to the approximation of accounting for
the effect of distortion through a multiplicative factor. From the
analysis of the hadron responses we have established the following:
\begin{itemize}
\item[i)] RPA correlations are strongly damped, but not
completely washed out: this is at variance with the polarized
($\vec{\rm p}$,$\vec{\rm n}$) experiment in Los Alamos at
$q\approx1.72$ fm$^{-1}$
\cite{mcclel92}, which seems to be compatible with free responses.
It remains to be understood which mechanism is responsible of the
complete suppression of the RPA correlations experimentally found;
\item[ii)] in contrast to (e,e$'$) scattering, the $q$-dependence
of the strength is well reproduced (of course, properly accounting
for the non-scalar-isoscalar channels in the (p,p$'$) reaction at
high momenta), confirming the validity of the assumption that
2p-2h contributions should be negligible for hadronic reactions;
\item[iii)] the height of the QEP depends on the value of
$\widetilde\sigma_{\rm tot}$, which is not well assessed: note
that if one uses the free value, 40 mb, the height of the (p,p$'$)
cross-section is well reproduced, whereas the height of the (p,n)
one is underestimated; \item[iv)] again in contrast to (e,e$'$)
scattering, the QEP comes out in the wrong position: the
calculations in $^{12}$C predict a hardening of the cross-section
in the whole range of momenta explored ($\sim1\div3$ fm$^{-1}$),
as it happens in the (p,n) but not in the (p,p$'$) reaction.
Items iii) and iv) together apparently point to the
need for contributions that appear only in (p,n) and not in (p,p$'$)
scattering;
\item[v)] the shift of the peak in the calculated responses is a
consequence of the distortion of the proton wave and it is {\it not}
the same in $^{12}$C and $^{40}$Ca, whereas the available data
appear to scale. Since the Glauber theory is equivalent to the
distorted wave Born approximation in the eikonal limit (which might
or might not be valid at the incident energy of 800 MeV), it would
be interesting to see if also in the DWBA framework a shell
dependence shows up.
Note also that the two-step response is not treated in a way fully
consistent with the Glauber theory, a fact that could have some
influence on the shape of the response functions;
\item[vi)] at high transferred momenta
the two-step term is sizable in the (p,n) reaction and,
because of the momentum dependence of the NN amplitudes, does not
give a smoothly increasing background, as it does at small
momenta, but rather a contribution peaked slightly below the QEP.
\end{itemize}
No final statement can be made, at the moment, on these issues:
however, we wish to stress the necessity of formulating a theory
of the nuclear response able to cope simultaneously with all the
different kinds of reactions. Accurate calculations are surely
needed, but only from a careful cross-referencing of the many
phenomena unveiled in the scattering on complex nuclei can one
hope to obtain a guide towards a solution of the difficult
many-body nuclear problem.

\acknowledgements

It is a pleasure to thank Prof. A. Molinari and Prof. S. Rosati
for their encouragement and for many helpful discussions.

\appendix{The transverse polarization propagator}
\label{app:A}

The transverse polarization propagator is defined as the trace
of the current-current propagator, namely
\begin{equation}
  \Pi_T({\bf q},{\bf q}';\omega) =
  \sum_{n}\Pi_{nn}({\bf q},{\bf q}';\omega),
\end{equation}
where $\Pi_{mn}$ is obtained by substituting
{\boldmath $(\sigma$}$\times\hat{\bf q})_m$ and
{\boldmath $(\sigma$}$\times\hat{\bf q})_n$ for the two vertex
operators in Eq.~(\ref{eq:piqq}) \cite{alb85};
$m$ and $n$ are spherical indices.

Introducing vector spherical harmonics, $\Pi_{mn}$ can be recast
in the following form:
\begin{equation}
  \Pi_{mn}({\bf q},{\bf q}';\omega) =
  \sum_{\scriptstyle J M\atop J_1 J_2}
  \Pi_{J;J_1 J_2}(q,q';\omega)
   Y^{(m)*}_{JJ_1M}(\hat{\bf q})\,Y^{(n)}_{JJ_2M}(\hat{\bf q}'),
\end{equation}
where
\begin{equation}
  \Pi_{J;J_1 J_2}(q,q';\omega) =
  {1\over2}\sum_{\ell\ell'}
     \left[{\hat\Pi}_J(q,q';\omega)\right]_{\ell\ell'}
     b^{J_1}_{J\ell}\,b^{J_2}_{J\ell'}.
  \label{eq:pij1j2}
\end{equation}
The quantities in (\ref{eq:pij1j2}) have been defined in
Sec.~\ref{subsec:polprop}. Using the addition theorem for the
vector spherical harmonics, namely
\begin{equation}
  4\pi\sum_{M=-J}^{J}\sum_{m}
  Y^{(m)*}_{JJ_1M}(\hat{\bf q})\,Y^{(m)}_{JJ_2M}(\hat{\bf q}') =
  \delta_{J_1J_2}(2J+1)
  P_{J_1}(\hat{\bf q}\cdot\hat{\bf q}'),
\end{equation}
one finds
\FL
\begin{eqnarray}
  \Pi_T({\bf q},{\bf q}';\omega) &=&
  \sum_{JJ'}{(2J+1)\over4\pi} \Pi_{J;J'J'}(q,q';\omega)
  P_{J'}(\hat{\bf q}\cdot\hat{\bf q}')\nonumber\\
  &=&  \sum_{JJ'M'} \Pi_{J;J'J'}(q,q';\omega)
    Y^*_{J'M'}(\hat{\bf q})Y_{J'M'}(\hat{\bf q}')
  {2J+1\over2J'+1},
\end{eqnarray}
i.~e., Eq.~(\ref{eq:pit}) with
$\Pi_{JJ'}(q,q';\omega)=\Pi_{J;J'J'}(q,q';\omega)$.

Using Eq.~(\ref{eq:pij1j2}) the diagonal part of the transverse
propagator can be written as
\begin{equation}
  \Pi_T({\bf q},{\bf q};\omega) = \sum_{J}{2J+1\over4\pi}
  \sum_{\ell\ell'}\left[{\hat\Pi}_J(q,q';\omega)\right]_{\ell\ell'}
  \sum_{J'} b^{J'}_{J\ell}\,b^{J'}_{J\ell'}.
\end{equation}
It is easy to check that $\sum_{J'} b^{J'}_{J\ell}\,b^{J'}_{J\ell'}
= \delta_{\ell\ell'}-a_{J\ell}a_{J\ell'}$ ($a_{J\ell}$ being
defined in Eq.~(\ref{eq:ajl})), from which the result of
Eq.~(\ref{eq:pijt}) follows immediately.

\appendix{The surface polarization propagator}
\label{app:B}

Here we briefly sketch the derivation of the longitudinal surface
polarization propagator (\ref{eq:pijls}). Similar calculations
apply for the transverse one.

As mentioned in Sec.~\ref{subsec:onestep}, the expression
(\ref{eq:olts}) for the surface vertex operator
$O^{\it surf}_{L}({\bf q},{\bf r})$ is oversimplified: actually,
in (\ref{eq:olts}) one should add, inside the momentum integral,
the matrix element of the probe spin longitudinal operator,
$<s_f|${\boldmath $\sigma$}$^{(p)}\cdot${\boldmath $\lambda$}$|s_i>$
(the treatment of the isospin is trivial).
Inserting the resulting expression in the formula
(\ref{eq:piqq}) for the polarization propagator, averaging over
the initial probe spin $s_i$ and summing over the final spin $s_f$
one finds (for simplicity, we consider only the diagonal part of
$\Pi_L$, i.~e. {\bf q}={\bf q}$'$)
\widetext
\begin{eqnarray}
  \Pi_L^{\it surf}({\bf q},{\bf q};\omega) &=&
  {1\over(2\pi)^4}{1\over|f_L(q)|^2}
  \int d{\bf b}\,d\mbox{\boldmath $\lambda$}
  \int d{\bf b}'\,d\mbox{\boldmath $\lambda$}'\,
  \hat{\mbox{\boldmath $\lambda$}}\cdot
  \hat{\mbox{\boldmath $\lambda$}}{}' \,
  f^*_L(\lambda) {\rm e}^{-i\chi^*_{\rm opt}(b)}
  {\rm e}^{-i({\bf q}-\mbox{\boldmath $\scriptstyle \lambda$})
  \cdot{\bf b}}
  \nonumber\\
  &&\times
  \Pi_L
  (\mbox{\boldmath $\lambda$},\mbox{\boldmath $\lambda$}';\omega)
  f_L(\lambda') {\rm e}^{i\chi_{\rm opt}(b')}
  {\rm e}^{i({\bf q}-\mbox{\boldmath $\scriptstyle \lambda$}')
  \cdot{\bf b}'}
  \nonumber\\
  &=& {1\over(2\pi)^4}{1\over|f_L(q)|^2}
  {4\pi\over3}\sum_{JM}\sum_{n}
  \int d{\bf b}\,d\mbox{\boldmath $\lambda$}
  \int d{\bf b}'\,d\mbox{\boldmath $\lambda$}'
  f^*_L(\lambda) {\rm e}^{-i\chi^*_{\rm opt}(b)}
  {\rm e}^{-i({\bf q}-\mbox{\boldmath $\scriptstyle \lambda$})
  \cdot{\bf b}}
  \nonumber\\
  &&\times \Pi_J(\lambda,\lambda';\omega)
  f_L(\lambda') {\rm e}^{i\chi_{\rm opt}(b')}
  {\rm e}^{i({\bf q}-\mbox{\boldmath $\scriptstyle \lambda$}')
  \cdot{\bf b}'}
  Y^*_{JM}(\hat{\mbox{\boldmath $\lambda$})}
  Y^*_{1n}(\hat{\mbox{\boldmath $\lambda$})}
  Y_{JM}(\hat{\mbox{\boldmath $\lambda$}}{}')
  Y_{1n}(\hat{\mbox{\boldmath $\lambda$}}{}'), \nonumber\\
  \label{eq:bpils}
\end{eqnarray}
\narrowtext
\noindent
where we have used the angular momentum expansion (\ref{eq:pil})
and set $\hat{\lambda}_n=
\sqrt{4\pi/3}Y_{1n}(\hat{\mbox{\boldmath $\lambda$}})$.
After some algebra, one can recast (\ref{eq:bpils}) in the
following form
\FL
\begin{eqnarray}
  &&\Pi_L^{\it surf}({\bf q},{\bf q};\omega) =
  {1\over|f_L(q)|^2} \sum_{J}\sum_{\ell m}
  {2J+1\over2\ell+1}a_{J\ell}^2 \nonumber\\
  &&  \times
  \int_0^\infty db\,b\, {\rm e}^{-i\chi^*_{\rm opt}(b)}
  \int_0^\infty db'\,b'\, {\rm e}^{i\chi_{\rm opt}(b')}
  \int_0^\infty d\lambda\,\lambda\, f^*_L(\lambda)
  \int_0^\infty d\lambda'\,\lambda'\, f_L(\lambda')
  \Pi_J(\lambda,\lambda';\omega) \nonumber\\
  &&  \times
  {\cal F}^*_{\ell m}(q;b,\lambda) {\cal F}_{\ell m}(q;b',\lambda'),
\end{eqnarray}
$a_{J\ell}$ being defined in Eq.~(\ref{eq:ajl}) and
\FL
\begin{equation}
  {\cal F}_{\ell m}(q;b,\lambda) = {1\over(2\pi)^2}
  \int_0^{2\pi} d\phi_b \int_0^{2\pi} d\phi_\lambda
  {\rm e}^{i({\bf q}-\mbox{\boldmath $\scriptstyle \lambda$})
  \cdot{\bf b}}
  Y_{\ell m}(\hat{\mbox{\boldmath $\lambda$}}).
\end{equation}
$\hat{\mbox{\boldmath $\lambda$}}$ is a versor in the ($x,y$)
plane, hence
\widetext
\FL
\begin{eqnarray}
  &&Y_{\ell m}(\hat{\mbox{\boldmath $\lambda$}}) \equiv
    Y_{\ell m}({\pi\over2},\phi_\lambda) \nonumber\\
  &&=
    (-1)^m\left[{(2\ell +1)(\ell-m)!\over
    4\pi(\ell+m)!}\right]^{1\over2}
    P^m_\ell(0){\rm e}^{im\phi_\lambda} \nonumber\\
  &&=  \left\{(-1)^m\left({2\ell+1\over4\pi}\right)^{1\over2}
    \left[{(\ell-m-1)!!(\ell+m-1)!!\over
    (\ell+m)!!(\ell-m)!!}\right]^{1\over2}
   \right.\nonumber\\&&\left.\phantom{=\big\{}\times
     {(-1)^{\ell+m}\over2}
     \left[(-1)^{(\ell+m)/2}+(-1)^{-(\ell+m)/2}\right]
     \right\} {\rm e}^{im\phi_\lambda} \nonumber\\
  &&=   K_{\ell m}{\rm e}^{im\phi_\lambda},
  \label{eq:ylm}
\end{eqnarray}
\narrowtext
\noindent
where $K_{\ell m}$ is defined as the the quantity in brackets in
(\ref{eq:ylm}).

Thus, employing a standard integral representation of the Bessel
function, one finds
\begin{equation}
  {\cal F}_{\ell m}(q;b,\lambda) =
  (-1)^m K_{\ell m} J_M(qb)J_M(\lambda b)
\end{equation}
and
\FL
\begin{eqnarray}
  && \Pi_L^{\it surf}({\bf q},{\bf q};\omega) =
    {1\over4\pi}\sum_{J}(2J+1) \nonumber\\
  &&  \times \int_0^\infty d\lambda\,\lambda\, f^*_L(\lambda)
    \int_0^\infty d\lambda'\,\lambda'\, f_L(\lambda')
    \Pi_J(\lambda,\lambda';\omega) \nonumber\\
  && \times \sum_{\ell m}c_{J\ell m}
     \int_0^\infty db\,b\, {\rm e}^{-i\chi^*_{\rm opt}(b)}
     J_m(qb)J_m(\lambda b)
     \int_0^\infty db'\,b'\, {\rm e}^{i\chi_{\rm opt}(b')}
     J_m(qb')J_m(\lambda' b'),
  \label{eq:cpils}
\end{eqnarray}
with $c_{J\ell m}$ given by Eq.~(\ref{eq:cjl}).

Using the well-known orthogonality relation
\begin{equation}
  \int_0^\infty db\,b\,J_m(qb)J_m(\lambda b)=
  {\delta(q-\lambda)\over\lambda}
\end{equation}
one can write
\FL
\begin{equation}
   \int_0^\infty db\,b\, {\rm e}^{i\chi_{\rm opt}(b)}
     J_m(qb)J_m(\lambda b) = {\delta(q-\lambda)\over\lambda}
   -g_m(\lambda,q),
  \label{eq:bgm}
\end{equation}
having defined $g_m(\lambda,q)$ in Eq.~(\ref{eq:gm}).
Substituting (\ref{eq:bgm}) in (\ref{eq:cpils}), it is then
straightforward to obtain Eq.~(\ref{eq:pijls}).

\figure{ph self-energy diagrams: (a) single particle diagrams;
  (b) interference diagrams.\label{fig1}}
\mediumtext
\figure{$\sigma\tau$ longitudinal and transverse response
  functions of $^{40}$Ca at $q=410$ MeV/c and $q=330$ MeV/c.
  Free (dashed line) and RPA (solid line)
  responses without spreading width and free (dotted line) and RPA
  (dot-dashed line) responses with spreading width are displayed.
  \label{fig2}}
\widetext
\figure{Structure function of $^{40}$Ca at $q=410$ MeV/c and $q=330$
  MeV/c. Free (dashed line) and RPA (solid line) contributions
  include the spreading width. Data are from ref.~\cite{meziani85}.
  \label{fig3}}
\mediumtext
\figure{$\sigma\tau$ longitudinal response functions of $^{12}$C
  and $^{40}$Ca at $q=1.54$ fm$^{-1}$ and $q=2.31$ fm$^{-1}$.
  Free (dashed line) and RPA (solid line) {\it volume} responses
  are reported, together with free (dotted lines) and RPA
  (dot-dashed lines) {\it surface} responses for two-values
  of $\widetilde\sigma_{\rm tot}$:
  $\widetilde\sigma_{\rm tot}=30$ mb (higher curves) and
  $\widetilde\sigma_{\rm tot}=40$ mb (lower curves).
  The spreading width is always included.
  The $^{40}$Ca surface responses have been multiplied by a factor 2.
  \label{fig4}}
\figure{As in Fig.~\ref{fig4}, but for $\sigma\tau$ transverse
  response functions. \label{fig5}}
\figure{$\sigma\tau$ surface longitudinal response functions
  (divided by $N_{\rm eff}$) of $^{12}$C and $^{40}$Ca at
  $q=1.54$ fm$^{-1}$ and $q=2.31$ fm$^{-1}$ with (SW) and without
  (no SW) inclusion of spreading width. In each plot the $^{12}$C
  free (dashed line) and RPA (solid line) responses are compared
  to the $^{40}$Ca free (dotted line) and RPA (dot-dashed line)
  responses. \label{fig6}}
\widetext
\figure{Two-step responses of $^{40}$Ca with gaussian amplitudes
  $|f(q)|^2=A\exp{(-\eta q^2)}$ at $q=1.4$ fm$^{-1}$ and
  $q=2.4$ fm$^{-1}$ for various values of $\eta$:
  (1) $\eta=0.001$, (2) $\eta=0.01$, (3) $\eta=0.1$,
  (4) $\eta=0.5$, (5) $\eta=1$;
  the arrows show the position of the QEP.
  The scale is arbitrary. \label{fig7}}
\figure{Inelastic (p,n) cross-sections on $^{12}$C
  at $q$ ranging from 1.16 fm$^{-1}$ to 2.31 fm$^{-1}$.
  The calculation based on Eq.~(\ref{eq:csa}) is reported
  (solid line), together with the separate one-step (dashed line)
  and two-step (dotted line) contributions and with the calculation
  based on Eq.~(\ref{eq:csb}) (dot-dashed line).
  Data are from the 795 MeV experiment of ref.~\cite{tadd91}.
  \label{fig8}}
\narrowtext
\figure{As in Fig.~\ref{fig8}, but for inelastic (p,p$'$)
  cross-sections on $^{12}$C at $q$ ranging from 1.42 fm$^{-1}$ to
  1.93 fm$^{-1}$. Data are from the 795 MeV experiment of
  ref.~\cite{chrien80}. \label{fig9}}
\end{document}